# Effect of assistive method on the sense of fulfillment with agency Modeling with flow and attribution theory

Dan Nanno and Hideyoshi Yanagisawa*

The University of Tokyo.
Tokyo, Japan
hide@mech.t.u-tokyo.ac.jp

**ABSTRACT**

Several assistive technologies for users' operations have been recently developed. A user's sense of agency (SoA) decreases with increasing system assistance, possibly resulting in a decrease in the user's sense of fulfillment. This study aims to provide a design guideline for an assistive method to maintain and improve the sense of fulfillment with SoA. We propose a mathematical model describing the mechanisms by which the assistive method affects SoA and SoA induces a sense of fulfillment. The experience in the flow state is assumed to be a sense of fulfillment. The assistance effect on the skill-challenge plane in flow theory is defined as an increase in skill and decrease in challenge. The factor that separates the two effects from attribution theory is the locus of causality, which is matched to the judgement of agency (JoA) from the two-step account of agency. We hypothesized that the assistance increases the perception of skill and sense of fulfillment is greater when the locus of causality is internal, rather than external. To verify this hypothesis, a game task experiment was conducted with assistance that varied with the ease of recognition. We hypothesized that a player's JoA is internal for hard-to-recognize assistance, resulting in a high sense of fulfillment. Experimental results supported this hypothesis.

**INTRODUCTION**

In recent years, various products have been digitized and made autonomous, thereby realizing miniaturization and high performance. These products considerably assist humans to easily obtain better results. However, the sense of agency (SoA) decreases for increasing assistance [1]. SoA is the sense that "I am the one who is causing or generating an action" [2]. We assumed that the SoA affect is not only a sense of control, but also a positive psychological experience such as a sense of fulfillment and intrinsic motivation, and that an increase in the assistance results in a decrease of the sense of fulfillment or intrinsic motivation. The main purpose of this research is to realize a design of an assistive method to maintain or improve a sense of fulfillment or intrinsic motivation. The problem remains that users feel that they do not control or decide the operation of the product [1], and research on the occurrence of SoA has been conducted [1-3]. The relationship between the assistive method and SoA has been presented in previous research [1]; however, the relationship between SoA and sense of fulfillment or intrinsic motivation is unclear. In this paper, we propose a mathematical model that explains the effect of SoA on the sense of fulfillment with assistance.

The proposed model describes in detail the mechanism by which SoA affects the sense of fulfillment. It is possible to use the model as a guideline for the design of an assistive method. The study was conducted as follows: 1) the model was developed using SoA, flow theory, and attribution theory; 2) the model was mathematically formulated; 3) a simulation was conducted to demonstrate the effect of SoA on the sense of fulfillment using the mathematical model; and 4) the hypothesis derived from the mathematical model was verified using a game task in which the assistance method was manipulated as a factor of SoA.

**THEORIES**

**Sense of Agency**

According to the comparator model [4], SoA occurs unconsciously upon comparison of the predicted and actual sensory feedbacks. However, the occurrence of SoA is related to the judgement of agency (JoA) because of action and other cues [5]. In the two-step account of agency [6], as shown in Figure 1, SoA is distinguished by two steps: the feeling of agency (FoA) described in the comparator model and JoA based on posterior reasoning. FoA is the non-conceptual, low-level feeling of being an agent of an action. Contrastingly, JoA is an explicit conceptual, interpretative judgement of being the agent. In addition, in cue integration theory [7,8], there are weights for each clue that are cues that affect SoA. The weight depends on the availability and reliability of cues.

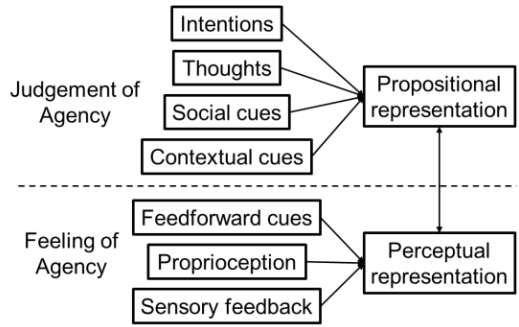

Figure 1. Two-step account of agency.

As a specific example of changing SoA, automated products decrease SoA [1]. The automation of a product is an increase in the amount of assistance. When the extent of assistance is large, sensory disagreement is likely to occur, and the assistance is noticed easily. Therefore, both FoA and JoA decrease.

**Flow Theory [9]**

Flow is a psychological state of fulfillment, concentration, controlled feeling, and intrinsic rewards. The flow state is prompted by the achievement of a goal, balance between skill and challenge, and appropriate feedback. Skill is the perceived ability to deal with a task, while challenge is the perceived difficulty of a task. The balance between skill and challenge occurs when both challenge and skill are balanced at the average level of the individual.

As shown in Figure 2, skill and challenge correspond to eight psychological states. The center of the figure represents the average challenge and skill levels of the individual. The greater the distance from the average level, the more strongly human beings experience each psychological state.

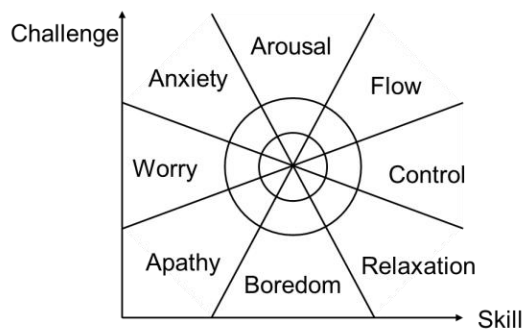

Figure 2. Classification of psychological state.

**Attribution Theory of Achievement Motivation**

In Weiner theory [10], the attribution of achievement vs. failure is classified by three dimensions: the locus of causality, stability, and controllability.

The locus of causality represents whether or not the cause of achievement or failure is internal or external. The internal cause is attributed to oneself, while the external cause is attributed to others. Stability refers to the variability; this is the expectation that the same result will be obtained in a future action. Controllability refers to whether or not the cause of achievement vs. failure is controllable, regardless of the locus of causality.

**MODELING**

**Definition**
We proposed a model by integrating several theories: the two-step account of agency, flow theory, and attribution theory. Before explaining the model, we define the concepts in this paper and organize the concepts from the theories.

Based on a two-step account of agency, we assumed that SoA consisted of FoA and JoA. Desantis [11] explained that the causal beliefs are an essential recondition for the intentional bindings effect to emerge. The intentional bindings effect is used as one of the indicators of SoA. The causal beliefs are seemingly a precondition for SoA. We assumed that the causal beliefs were equal to the locus of causality.

In the flow state, the sense of fulfillment is obtained when the desire to experience the flow state is fulfilled. In this paper, the sense of fulfillment is the positive psychological experience obtained by fulfilling the desire for a more advanced flow state.

**Qualitative Model**
Figure 3 shows the outline of the proposed model. We created the model by integrating SoA, attribution theory, and flow theory. The assistance changed the skill and challenge; the skill increased while the challenge decreased. Based on the two-step account of agency, SoA is distinguished by FoA and JoA. Based on attribution theory, the locus of causality affected the amount of change in both the skill and challenge. We assumed that the locus of causality is the JoA. Based on flow theory, both the skill and challenge affected the sense of fulfillment. Based on the model, the sense of fulfillment changed according to SoA.

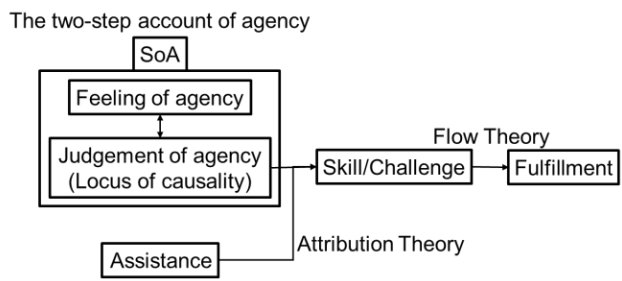

Figure 3. Overview of model.

**Transitions to Flow State via Assistance**
Figure 4 shows an overview of transitions to the flow state via two types of assistance. The origin (α state) in Figure 4 represents the state where both the skill and challenge are perceived as balanced at the average level. That is, the center in Figure 2 also represents the origin in Figure 4. The average state is the expectation of the skill and challenge for oneself. The origin changed along with the individual and action. Human beings perceive skill and challenge as high when both the skill and challenge are perceived to be greater than average.

In the flow state, the skill and challenge were balanced and the task was achievable. For the unachievable task, the challenge was perceived to be greater than the challenge in the average state. The β state in Figure 4 represents the state perceived with the unachievable task. The two γ states shown in Figure 4 are examples of transitions from the β state with two assistances. In the γ1 state, the distance from the α state was small; in the γ2 state, it was large. For the transition from the β state to the two γ states, the perceived skill increased and the perceived challenge decreased. We expected an increase in the perceived skill when humans perceived assistance to strengthen their ability, and that the perceived challenge would decrease when humans perceived assistance, lessening the task difficulty. The γ1 state represents a state where assistance strongly influenced the decrease of the perceived challenge, and the γ2 state represents a state where assistance strongly influenced the increase of the perceived skill. We compared these two types of assistances. The distance from the α state to the γ states was greater to γ2 than to γ1, resulting in a stronger flow state. As a sense of fulfillment is experienced in the flow state, the sense of fulfillment was greater in the γ2 state than that in the γ1 state.

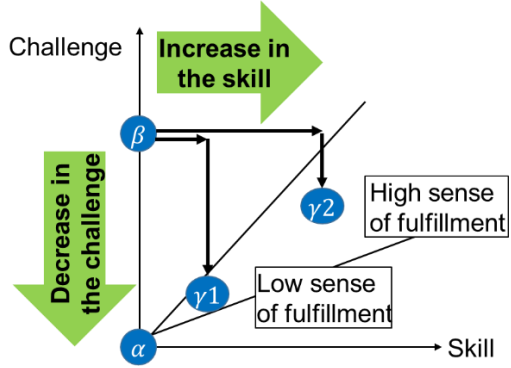

Figure 4. Transition to flow state via two types of assistance.

**Classification of Assistance via Attribution Theory**

Using attribution theory, we investigated the factors that classify whether the perceived skill increases or the perceived challenge decreases. The human beings perceived increases in the skill as causes of achievement when it was the self that caused the achievement. In this case, the locus of causality was internal. The human beings perceived decreases in the challenge as causes of the achievement when it was others that caused the achievement. For this case, the locus of causality was external. The locus of causality classified the cause of achievement with assistance as an increase in the perceived skill or a decrease in the perceived challenge.

When the locus of causality is internal, it is assessed by the result of the self-caused action. We assumed that the locus of causality is a judgment based on posterior reasoning. SoA is distinguished from FoA based on sensory feedback and from JoA based on posterior reasoning. We treated the locus of causality as JoA.

Based on the above information, we hypothesized that as the assistance increases, the perception of skill and sense of fulfillment is higher when the locus of causality is internal rather than external.

**Mathematical Model**

**Perception of the skill and the challenge**

Prior to executing a task, humans have expectations of their own skill level and the challenge of the task (the average state). We assumed that these expectations could be represented by a normal distribution with expected values and variance, called a prior distribution. We assumed that the likelihood of actual skill and actual challenge could be represented by a normal distribution with expected values and variance, called the likelihood function. From Bayesian inference [12], the posterior distribution is calculated from the prior distribution and likelihood function. The expected value of the posterior distribution represents the perceived value. From Bayesian inference, the expected value of the posterior distribution, $\eta$, is expressed by equation (1).

$$\eta = \frac{\mu v + \varepsilon N}{v + N} \quad (1)$$

Here, $\mu$ is the expected value of the prior distribution, $v$ is the variance of the prior distribution (uncertainty), $\varepsilon$ is the expected value of the likelihood function, and $N$ is the variance of the likelihood function (noise).

We represented the perception of skill and challenge with $S$ and $C$, respectively. Because the average state was the origin, the expected value of the prior distribution was 0. Thus, equations (2) and (3) represent the perception of the skill and challenge.

$$S = \frac{N_s \delta_s}{v_s + N_s} \tag{2}$$

$$C = \frac{N_c \delta_c}{v_c + N_c} \tag{3}$$

Here, $\delta$ is the prediction error that is the difference between the expected value of the prior distribution and expected value of the likelihood function; and subscripts $s$ and $c$ represent the skill and the challenge, respectively.

We defined the strength of the psychological state, $H$, as the distance from the origin (the average state), and is given by

$$H = (S^2 + C^2)^{\frac{1}{2}} \tag{4}$$

The region of the flow state was sandwiched between two straight lines, with slopes $t_{high}$ and $t_{low}$, through the origin. In this region, the strength of the psychological state was sufficiently large. We defined the flow state as when equations (5) and (6) held. In the flow state, we assumed that $H$ corresponded to the degree of the sense of fulfillment.

$$t_{low} \leq \frac{C}{S} \leq t_{high} \tag{5}$$

$$H \gg 0 \tag{6}$$

**State transitions in the Skill-Challenge space via assistance**

We formalized the prediction error of each state shown in Figure 4. The α state was the average state. Equation (7) represents the prediction error of the skill and challenge.

$$\delta_s(\alpha) = \delta_c(\alpha) = 0 \tag{7}$$

Here, α represents the α state. The actual challenge in the β state was sufficiently higher than the α state, and the actual skill was the same as in the α state. Equations (8) and (9) represent the prediction error of the skill and challenge of the β state.

$$\delta_s(\beta) = 0 \tag{8}$$

$$\delta_c(\beta) \gg 0 \tag{9}$$

Here, β represents the β state; and the assistance increases the skill and decreases the challenge. In the γ states, the assistance changed the prediction error of the skill and challenge. The total amount of prediction errors that changed due to the assistance is represented by $x$; $x$ was greater than $\delta_c(\beta)$ because the task became achievable in the γ state. $x$ consists of $x_s$ (the change in the prediction error of the skill) and $x_c$ (the change in the prediction error of the challenge). We assumed that the total change of the prediction error was constant when the amount of assistance was constant. Equations (10) and (11) represent the prediction error of the γ states.

$$\delta_s(\gamma) = \delta_s(\beta) + x_s \tag{10}$$

$$\delta_c(\gamma) = \delta_c(\beta) - x_c \tag{11}$$

γ represents the γ states.

**Attribution theory**

We defined the locus of causality as a continuous quantity $L$ of value 0 to 1; equal to one when the location of the cause was completely within the self and zero when it was completely within the other. In other words, the larger the value of $L$, the stronger the judgment that oneself caused the action and result. The locus of causality classified the increase in the perceived skill and decrease in the perceived challenge. The value of $x_s$ was larger when L was larger. Also, $x_c$ was larger when $L$ was smaller. We represented this relation as a first-order relation with equations (12) and (13).

$$x_s = x \cdot L \tag{12}$$

$$x_c = x \cdot (1 - L) \tag{13}$$

**Relation of SoA and the locus of causality**

We formalize SoA as a summation of FoA and JoA, and is given by

$$SoA = FoA + JoA. \qquad (14)$$

We assumed that JoA corresponds to the locus of causality:

$$L = JoA. \qquad (15)$$

**Function model of the sense of fulfillment**

From equations (16)–(19), the sense of fulfillment is represented by the transition to the flow state via the assistance.

$$H = (A_1 \cdot JoA^2 + A_2 \cdot JoA + A_3)^{\frac{1}{2}} \qquad (16)$$

$$A_1 = x^2 \left( \left( \frac{N_s}{\nu_s + N_s} \right)^2 + \left( \frac{N_c}{\nu_c + N_c} \right)^2 \right) \qquad (17)$$

$$A_2 = 2x \left( \frac{N_c}{\nu_c + N_c} \right)^2 (\delta_c(\beta) - x) \qquad (18)$$

$$A_3 = \left( \frac{N_c}{\nu_c + N_c} \right)^2 (\delta_c(\beta) - x)^2 \qquad (19)$$

$A_{1\sim3}$ is a variable for simplifying the expression.

**Simulation using the mathematical model**

Figure 5 shows plots of the α, β, and γ states for JoA values of 0.1, 0.5, and 0.9 when the uncertainty and noise in each state were constant. As shown, in the γ state, the greater the value of JoA, the larger the distance from the α state.

Regarding equations (16)–(19), when $x$, $N_s$, $\nu_s$, $N_c$, $\nu_c$, and $\delta_c(\beta)$ were constant, $A_{1\sim3}$ was also constant and exceeded zero. $H$ monotonically increased with respect to JoA. We hypothesized that the sense of fulfillment would be high when the JoA was internal. Figure 5 demonstrated this relation.

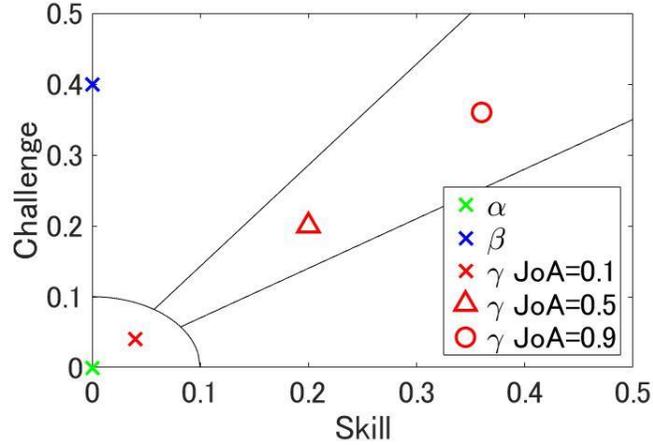

Figure 5. Simulation result of the trend in the challenge and skill space for varied JoAs.

**EXPERIMENT**

The hypothesis obtained from the mathematical model was that the sense of fulfillment was high when JoA was internal. To verify this hypothesis, we proposed a novel experimental method. The method was used to evaluate the strength of the flow state between two conditions in which JoA was changed by varying the ease of recognition of the assistance.

**Method**
    We assumed that a hard-to-recognize assistance produced an internal JoA, while an easy-to-recognize assistance produced an external JoA. We used the flow score scale as an indicator of a sense of fulfillment in the flow state [13]. We used a shooting game with the above-mentioned two types of assistance as an operation task in the experiment.
    In this experiment, subjective evaluation, physiological data, and behavioral data were obtained as responses from participants. Subjective evaluations were performed for each experimental condition. We used the flow state scale for the occupational tasks [13] to evaluate the strength of the flow state. We used the causal dimension scale [14] and direct questions of the cause attribution to evaluate the locus of causality. Furthermore, we created scales to assess SoA, increase in skill, decrease in challenge, and recognition of assistance. We used eye blink frequency using JINS MEME as a physiological index of concentration. We used voluntary selection as a behavioral indicator of preference.

**Task**
    We adopted a shooting game as a task. Figure 6 shows the outline of an experimental trial. Players shot upward from the player's icon and aimed at the target icon. The player icon moved from the lower left to the lower right at a constant speed, and the target icon was stopped at the upper side. When the player clicked, after 250 ms, the drawing of the shooting and hit or miss determination are presented to the player as feedback. A player clicked "again" to proceed to the next trial.
    We evaluated each player's skill using the difference between the center of the target and location hit. We evaluated the difficulty of the task using the width of the target. When there was no assistance, the width of the target was the same as the width of the area hitting the target. The assistance was performed by increasing the width of this area, and this increase was set as the assistance amount. The position of the player at the time of clicking was classified as either the area hitting the target without assistance, the area hitting with assistance, or the missing area.
    Figure 7 shows the two assistance methods with varying ease of recognition. For the hard-to-recognize assistance condition, the speed of movement of the player's icon from clicking to shooting changed continuously, and the player's icon was shot directly above. For the easy-to-recognize assistance, the player's icon moved at a constant speed from clicking to shooting and shot in the direction towards the target.
    Players could visually recognize the easy-to-recognize assistance at the time of shooting. By contrast, the players could hardly recognize the hard-to-recognize assistance visually at the time of shooting.

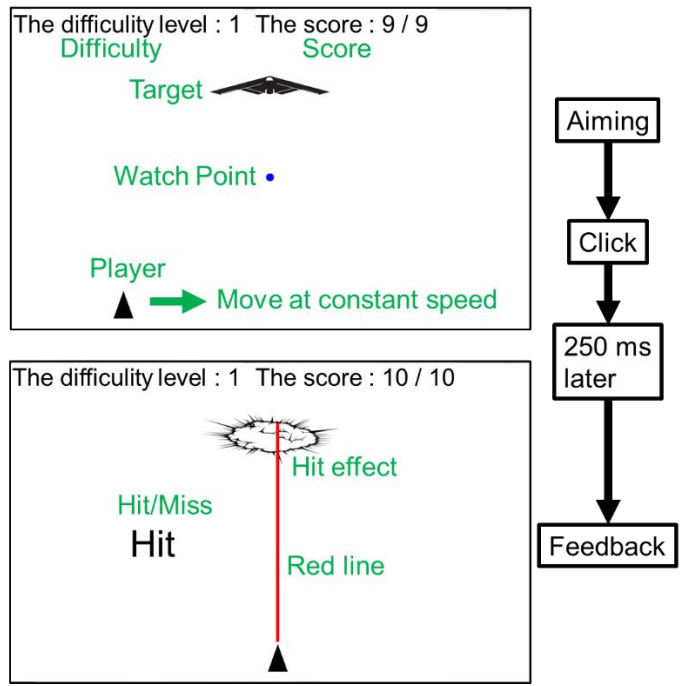

Figure 6. Experimental task trial.

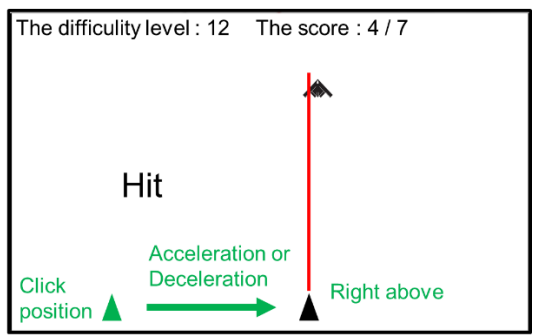

(a)

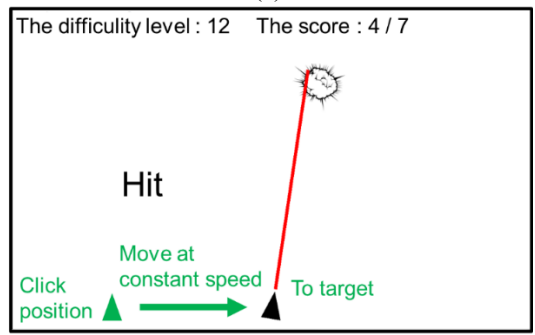

(b)

Figure 7. Two assistance methods with different ease of cognitions, (a) Hard-to-recognize assistance and (b) Easy-to-recognize assistance.

**Participants**

Eleven Japanese, right-handed, healthy male and female volunteers between 22 and 24 years old with normal or corrected-to-normal vision, and normal hearing participated in this study.

**Procedure**

The participants played five different sessions: practice, difficulty adjustment, high difficulty, assisted, and free. During the practice session, the subjects performed the task until they were familiar with it. During the difficulty adjustment session, the difficulty levels at which the subjects' skill and challenge levels were balanced were determined. The target width was dynamically changed in each trial using the standard deviation of the distance between the center of the target and hit position as an indicator. We looked for a target width that roughly matched the standard deviation of the difference. We assumed that subjects were in the α state in Figure 4 at the sought-after difficulty level.

In the high difficulty session, participants performed a task at an obviously higher difficulty level compared to the desired difficulty level. We set the width of the target for the high difficulty level to 25% of the searched width of the target. We assumed that subjects were in the β state in Figure 4 at the high difficulty level.

During the assisted session, the participants performed the task at a high difficulty level with assistance. At the end of the session, the participants responded to the questionnaire for the subjective evaluations. During the free session, the participants freely selected two assistants and performed the tasks as many times as desired.

**Analysis**

We compared the participants' responses between the easy-to-recognize and the hard-to-recognize assistances using a t-test.

**Results**

The differences between the average flow scores of the participants in the easy-to-recognize assistance vs. the hard-to-recognize assistance were not significant ($p = 0.244$). Focusing on the data of each participant, two participants responded that the value of the locus of causality was higher in the easy-to-recognize assistance than in the hard-to-recognize assistance. We classified the two experimental conditions of each participant into the internal and external JoA conditions using the subjective scores of the locus of causality.

Figure 8 shows a comparison of the flow scores in the internal and external JoA conditions. The flow scores were significantly higher in the internal JoA condition compared with those in the external JoA condition ($p = 0.008$).

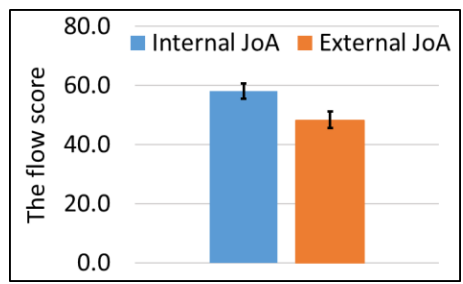

Figure 8. Comparison of the average flow score between the internal and external JoAs.

## DISCUSSION

The model proposed in this paper describes the effect of SoA on the sense of fulfillment. According to equations (16)–(19), JoA, $x$, $N_s$, $v_s$, $N_c$, $v_c$, and $\delta_c(\beta)$ changed the strength of the psychological state H. That is, the sense of fulfillment in the flow state was changed by JoA, the expectation of skill and challenge, the actual skill and challenge, and the amount of assistance. H monotonically increased with respect to the JoA when $x$, $N_s$, $v_s$, $N_c$, $v_c$, and $\delta_c(\beta)$ were constant. That is, the sense of fulfillment in the flow state increased when JoA was internal.

According to Figure 5, with the same amount of assistance, the same β state transitioned to different γ states. This indicated the possibility that different psychological states were obtained when the assistive method was different, even if the same amount of assistance was used.

We discussed the validity of the model by verifying our hypothesis that the sense of fulfillment should increase when JoA is internal. We considered the correspondence with the proposed model of the concepts in self-determination theory [15]. Self-determination theory explains the relationship between autonomy and psychological health. Autonomy is necessary for intrinsic motivation, and fulfilling the needs for autonomy provides humans with the intrinsic motivation and psychological health. The need of autonomy is the desire to control and determine the behavior of the self by the self. When humans feel autonomy with an action, they judge the action caused by themselves. We believe that JoA is the autonomy of action. In the self-determination theory, the feeling of autonomy is necessary for intrinsic motivation, and psychological health is obtained with intrinsic motivation. Meanwhile, in the model, the sense of fulfillment is obtained by fulfilling the desire for a more advanced flow state. In the flow state, humans experience the feeling of intrinsic rewards. When humans feel intrinsically rewarded, they are intrinsically motivated. Humans are intrinsically motivated by the desire for the flow state and feel the sense of fulfillment. The psychological health and sense of fulfillment are similar in that they are positive psychological experiences associated with intrinsic motivation. From the aforementioned discussion, the proposed model explains the relationship described by self-determination theory. Self-determination theory supports our hypothesis.

We conducted an experiment to verify our hypothesis. As a result of the experiment using 11 subjects, the t-test of the flow scores of the internal and external JoA conditions showed significant differences with a significance level of 0.01. These experimental results supported our hypothesis.

The subjective score of the 'self-ability cause' was significantly higher in the internal JoA condition than that in the external JoA condition ($p = 0.002$). In addition, the subjective score of the 'assistance cause' was significantly higher in the internal JoA condition than in the external JoA condition ($p = 0.091$). Figure 9 shows the subjective scores of the attribution cause for the internal and external JoA conditions.

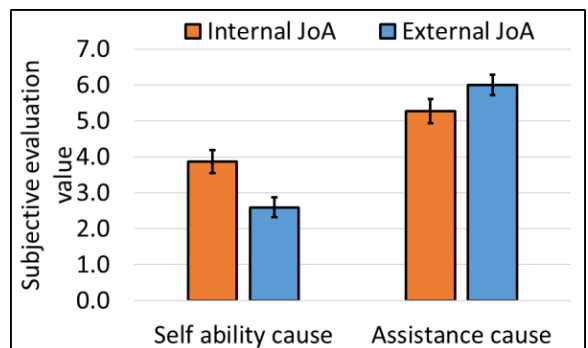

Figure 9. Average score of the attribution cause.

Figure 4 demonstrates that the perceived skill and challenge increased and decreased, respectively, in the transition from the β state to the two γ states. The β state corresponded to the high difficulty session in the experiment, while the two γ states corresponded to the assisted session in the experiment. Figure 10 shows the variations in the skill and challenge. The subjective score of 'the increase in the skill' was significantly higher in the internal JoA condition than that in the external JoA condition ($p = 0.019$). In addition, the difference in the subjective scores of the internal and external JoA conditions was not significant ($p = 0.246$).

To confirm that the increase in skill caused the change in JoA, we performed a t-test of the indicator of the actual skill. In the experiment, we recorded the score of the task and difference between the center of the target and hit position as an indicator of the actual skill. The difference between the scores of the task between the internal and external JoA conditions was not significant ($p=0.303$). In addition, the difference in the distances of the center of the target and hit position for the internal and external JoA conditions was not significant (*p*=0.267). From the aforementioned discussion, the sense of increased self-ability was changed by JoA, even though the actual skill did not vary.

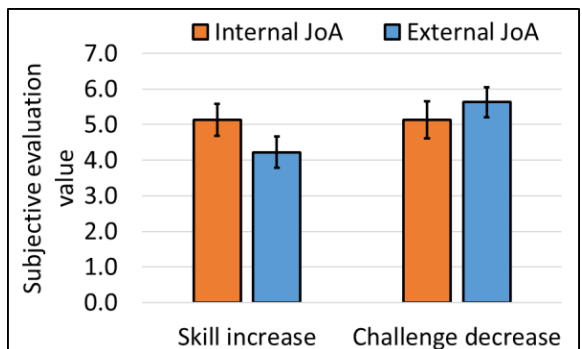

Figure 10. Average score of the changes in skill and challenge.

## CONCLUSION

In this paper, we proposed a mathematical model describing the effect of SoA on the sense of fulfillment when an unachievable task becomes achievable via assistance. We developed the mathematical model using SoA, flow theory, and attribution theory. Figure 3 and equations (16)–(19) represent the proposed model.

By proposing the model, we described and mathematically explained the effect of SoA in the product operation on the sense of fulfillment. Based on the model, JoA, the expectation of skill and challenge, actual skill and challenge, and amount of assistance changed people's psychological states, which transitioned via the assistance and sense of fulfillment in the flow state.

We confirmed the validity of the mathematical model using the following steps, although the number of subjects used in the experiment was not sufficiently large to verify the hypothesis.
1) The model explained the relationship between autonomy and sense of fulfillment, as explained in self-determination theory, as the relationship between a higher sense of fulfillment and internal JoA.
2) We verified our hypothesis that the sense of fulfillment is high when JoA is internal. The experimental results supported our hypothesis obtained from the model.
3) The experimental results showed that the assistance increased the perceived skill and decreased the perceived challenge. These results supported the model.

Experimental verification of the mathematical model made it possible to use the model to simulate the psychological states and sense of fulfillment. It is also possible to use the model as a guideline for the design of the assistive method.

Because this model is at a basic level, its practical use is assumed. For the UX design of the device, the model can be used to design the assistive method, in an entertaining way. For industrial equipment, it is possible to design an assistive method of the equipment that is operated without losing the intrinsic motivation or sense of responsibility, even if there are many automatic operations. For education, it is possible to design learning assistance teaching materials that enhance learning motivation.

## ACKNOWLEDGEMENTS

This research was carried out through a collaborative research project with the Sony Corporation. We would like to express our gratitude to Dr. Kazuko Yamagishi of Sony Global Manufacturing & Operations Corporation and to those who cooperated.

We would also like to express our gratitude to Professor Tamotsu Murakami of the University of Tokyo, Dr. Kazuki Ueda of the University of Tokyo, and to the members of the design engineering laboratory of the University of Tokyo.